\documentclass[onecolumn,preprintnumbers,amsmath,amssymb,superscriptaddress]{nature}
\usepackage{amssymb}
\usepackage{mathrsfs}
\usepackage{graphicx}
\usepackage{dcolumn}
\usepackage{bm}
\usepackage{amsmath}
\usepackage{amsfonts}
\usepackage{color}
\usepackage{float}
\title{Negative differential resistance and characteristic nonlinear electromagnetic response of a  Topological Insulator}

\author{Ching Hua Lee$^{2,*}$ \& Xiao Zhang$^{1,*}$ \& Bochen Guan$^1$}

\begin{document}

\maketitle

\begin{affiliations}
 \item Department of Physics, Sun-Yat-Sen University, Guangzhou, China.
 \item Department of Physics, Stanford University, CA 94305, USA. 
*These authors contributed equally to this work. Correspondence and requests for materials should be addressed to Xiao Zhang (email: yngweiz@gmail.com).
\end{affiliations}

\begin{abstract}
Materials exhibiting negative differential resistance have important applications in technologies involving microwave generation, which range from motion sensing to radio astronomy. Despite their usefulness, there has been few physical mechanisms giving rise to materials with such properties, i.e. GaAs employed in the Gunn diode. In this work, we show that negative differential resistance also generically arise in Dirac ring systems, an example of which has been experimentally observed in the surface states of Topological Insulators. This novel realization of negative differential resistance is based on a completely different physical mechanism from that of the Gunn effect, relying on the characteristic non-monotonicity of the response curve that remains robust in the presence of nonzero temperature, chemical potential, mass gap and impurity scattering. As such, it opens up new possibilities for engineering applications, such as frequency upconversion devices % with strong frequency multiplication properties,
which are highly sought for terahertz signal generation. Our results may be tested with thin films of Bi$_2$Se$_3$ Topological Insulators, and are expected to hold qualitatively even in the absence of a strictly linear Dirac dispersion, as will be the case in more generic samples of Bi$_2$Se$_3$ and other materials with
topologically nontrivial Fermi sea regions.
\end{abstract}

\date{\today}

\maketitle
%\tableofcontents

\section{Introduction}

%To address the highly desirable need for efficient frequency
%multiplication to higher harmonics, Mikhailov et al.$^{5}$ proposed
%Graphene as a new material frequency multiplication, owing to its
%unique linear Dirac dispersion $^{6-10}$. These predictions were subsequently supported experimentally$^{11}$. 
  Topological Insulators (TIs) are a new class of
materials with a fully insulating gap in the bulk but gapless
(conducting) Dirac fermion states on the surface$^{1-3}$, and have garnered tremendous interest in
condensed-matter physics, material science and electrical
engineering communities$^{1-14}$. Recent
experimental realizations of TI states in compounds like
${\text{Hg}}{\text{Te}}$, ${\text{Bi}}_{1-x}{\text{Sb}}_{x}$ and
${\text{Bi}}_{2}{\text{Te}}_{3}$ fueled further enthusiasm in their
possible application in devices. The Dirac cones on the surface of 3-dimensional TIs are reminiscent of the Dirac cones in 2-dimensional Graphene, another exotic material which has attracted considerable attention$^{15-21}$. Notably, Graphene has been theoretically predicted$^{19-20}$ and subsequently experimentally shown$^{21}$ to exhibit strong nonlinear electromagnetic response owing to its unique linear Dirac dispersion $^{15-21}$. 

%It is thus natural to explore the potential applications of TIs in analogy to the of the nonlinear response of Graphene$^{6}$. 
Inspired by these properties of Graphene, we ask if similar, if not more desirable, nonlinear behavior is also present in the TIs. As we will show in this work, the answer is in the affirmative: In fact, the response curve of TIs is even more nonlinear, with an exotic regime of negative differential resistance that persists even in the absence of a strictly linear dispersion. More precisely, the response curve takes the form of an 'N' shape with negative differential resistance in the middle segment, similar to the shape of the response curve leading to the Gunn effect in GaAs$^{22}$, but due to a completely different physical mechanism. This suggests an array of potential optoelectronics applications beyond those of Graphene and the Gunn diode.

%are band insulators with band inversion driven by spin-orbit coupling$^{5}$ and interaction$^{6}$.
The enhanced nonlinearity of the response of TI surface states can be understood as follows. A TI heterostructure has two conducting surfaces, the top and bottom surfaces, while a Graphene sheet only has one. Due to the existence of the substrate in the TI heterostructure, structural inversion
symmetry has to be broken, leading to a breaking of the degeneracy of the
two TI surface states which opens up a Rashba-type spin
splitting$^{23}$. This results in an unique Dirac ring bandstructure
which exhibits a much stronger nonlinear electromagnetic response
than a Dirac cone alone, thereby opening up a venue for interesting
physics as well as potential applications. 

%The window of electromagnetic frequencies from 0.3 to 20 THz
%(also known as Terahertz gap) has been a very active area of research across a multitude of disciplines in engineering, material science and medicine$^{1-3,5}$. But despite their importance, inexpensive and compact sources for THz radiation are still lacking. Currently, low-power electromagnetic radiation within the Terahertz gap are most commonly produced via nonlinear multiplication
%(upconversion) of lower-frequency radiation$^{3-5}$. This, as often achieved with GaAs Schottky barrier diodes, have high upconversion efficiency for the first few harmonics but poor efficiencies for the higher order ones$^{3-5}$. In
%this work, we shall study the electromagnetic response of Dirac ring
%systems, and show that they provide a novel way for frequency
%upconversion that is more efficient than that of conventional
%technology.

In this work, we model a TI heterostructure as a Dirac ring system, and analytically and numerically study its semiclassical nonlinear response. We first consider the case of an ideal Dirac ring, i.e. at zero mass gap, temperature and
impurities. Next we consider deviations from these ideal conditions,
and crucially show that the characteristic features of the response curve remain
robust. We further discuss how this semiclassical analysis can be
generalized to a more general setting with scattering and/or
Rashba-like dispersion, and its implications for the output signal.
Finally, we discuss some experimental proposals and engineering
applications.
%We develop a semiclassical approach for the non-linear response in a material with generic dispersion. It works for arbitrarily large external field amplitudes and scattering times. Exact analytic results are obtained for generic driving fields, with a particularly elegant expression for the sinusoidal case. We analytically and numerically study the effects of temperature, scattering and mass gap for a Dirac cone, as well as its broadening into a ring (any better name?)

%We first introduce some basic results on the semi-classical electromagnetic response of a generic Hamiltonian. With that, we present our main result on the characteristic nonlinear response curve of an ideal Dirac ring system. The qualitative shape of this curve will be shown to remain .......
\section{Results}
We first introduce some basic theory on the semi-classical electromagnetic response of a generic Hamiltonian. With that, we present our main results on the characteristic nonlinear response curve of Dirac ring systems. Such systems have been detected in the surface states of thin films of ${\text{Bi}}_{2}{\text{Se}}_{3}$ TIs via ARPES experiments$^{23}$, and we will return to discussing the experimental signatures of our results after developing its theory.

\subsection{Theory of semiclassical response}

Consider a generic system described by a Hamiltonian $H(\vec p)$ under the influence of a driving field $\vec E(t)$. At the semi-classical level, the field shifts the crystal momenta $\vec p$ of the partially occupied bands, leading to an induced current
\begin{eqnarray}
\vec J(t)&=&e\langle \vec v\rangle_t= \frac{e}{(2\pi \hbar)^2}\int d^2 \vec p g(\vec p,t)\vec v
%&=& e\int d^2 \vec p F_0(\vec p -e \vec A_{eff}(t))\nabla_{\vec p}\epsilon(p) \notag\\
%&=& e\int d^2 \vec p F_0(\epsilon(p))\nabla_{\vec p}\epsilon(\vec p+\vec p_0)
\end{eqnarray}
where $\vec v=\nabla_{\vec p}\epsilon(\vec p)$ is the canonical velocity and $\epsilon(\vec p)$ is the eigenenergy for a particular band. In other words, $\vec J$ is the expectation value of the current $e\vec v$ over states weighted by the time-dependent equilibrium occupation function $g(\vec p ,t )$. %the induced current $\vec J$ is given by the expectation value of the current $e\vec v$ over all occupied states,  For each partially occupied band, the induced current $\vec J$ is given by%is the  across states with all values of momentum $\vec p$, weighted by the time-dependent occupation function $g(\vec p ,t )$:
As will be shown rigorously in Sect. \ref{sect:boltzmann2}, the equilibrium occupation function takes the form of the Fermi-Dirac distribution $g(\vec p,t)=F_0(\vec p-\vec p_0(t))$, but with momentum shifted by an \emph{effective} driving impulse $\vec p_0(t)=e\int^t\vec E_{eff}(t')dt'$.  We shall elaborate on exactly how $\vec E_{eff}$ depends on $\vec E$ and the scattering time $\tau$ in Sect. \ref{sect:boltzmann}. For now, we shall proceed by treating $\vec p_0$ as an external influence, and note that $\vec E_{eff}=\vec E$ in the limit of zero scattering. This approach considers only intra-band transport processes, and is valid when $\vec p_0$ originates from an oscillatory electric field with $\hbar\Omega\ll \text{max}\{\mu,T,m\}$, i.e. with frequencies under $20$ THz in typical applications. Henceforth, we shall work in units where $2\pi\hbar=1$ and $k_B=1$ for notational simplicity.
%($\tau \rightarrow \infty$).
%$\vec p \rightarrow \vec p +e \int^t\vec E(t') dt'=\vec p - e $ to the  is taken to
%where the velocity operator is $\vec v =\nabla_{\vec p}\epsilon$ and we have defined $\vec p_0=e\vec A_{eff}(t)$ for ease of notation.

If the bands of the hamiltonian are isotropic with eigenenergies $\epsilon(\vec p)=\epsilon(p)$ where $p=|\vec p|$, $F_0$ depends only on $p$ via $\epsilon(p)$ and we can further express $\vec J$ as
\begin{eqnarray}
\vec J &=& e\int d^2 \vec p F_0(\epsilon(\vec p-\vec p_0))\nabla_{\vec p}\epsilon(\vec p) \notag\\
&=& e\int d^2 \vec p F_0(\epsilon(\vec p))\nabla_{\vec p}\epsilon(\vec p+\vec p_0) \notag\\
%&=&e\int F_0(\epsilon(p))\left[\int \nabla_{\vec p}\epsilon\left(\vec p + \vec p_0\right) d\theta \right]pdp\notag\\
&=& e\int \vec j(p)dp
\label{J}
\end{eqnarray}
where
\begin{equation}
\vec j(p)=pF_0(\epsilon(p))\int \nabla_{\vec p}\epsilon\left(\vec p + \vec p_0\right) d\theta
\label{jp}
\end{equation}
is the contribution from the $p$-momentum shell. This decomposition is particularly useful since $\vec j(p)$ can often be expressed in closed-form, although $\vec J$ usually cannot be.

\begin{figure}[centered]
\includegraphics[scale=.32]{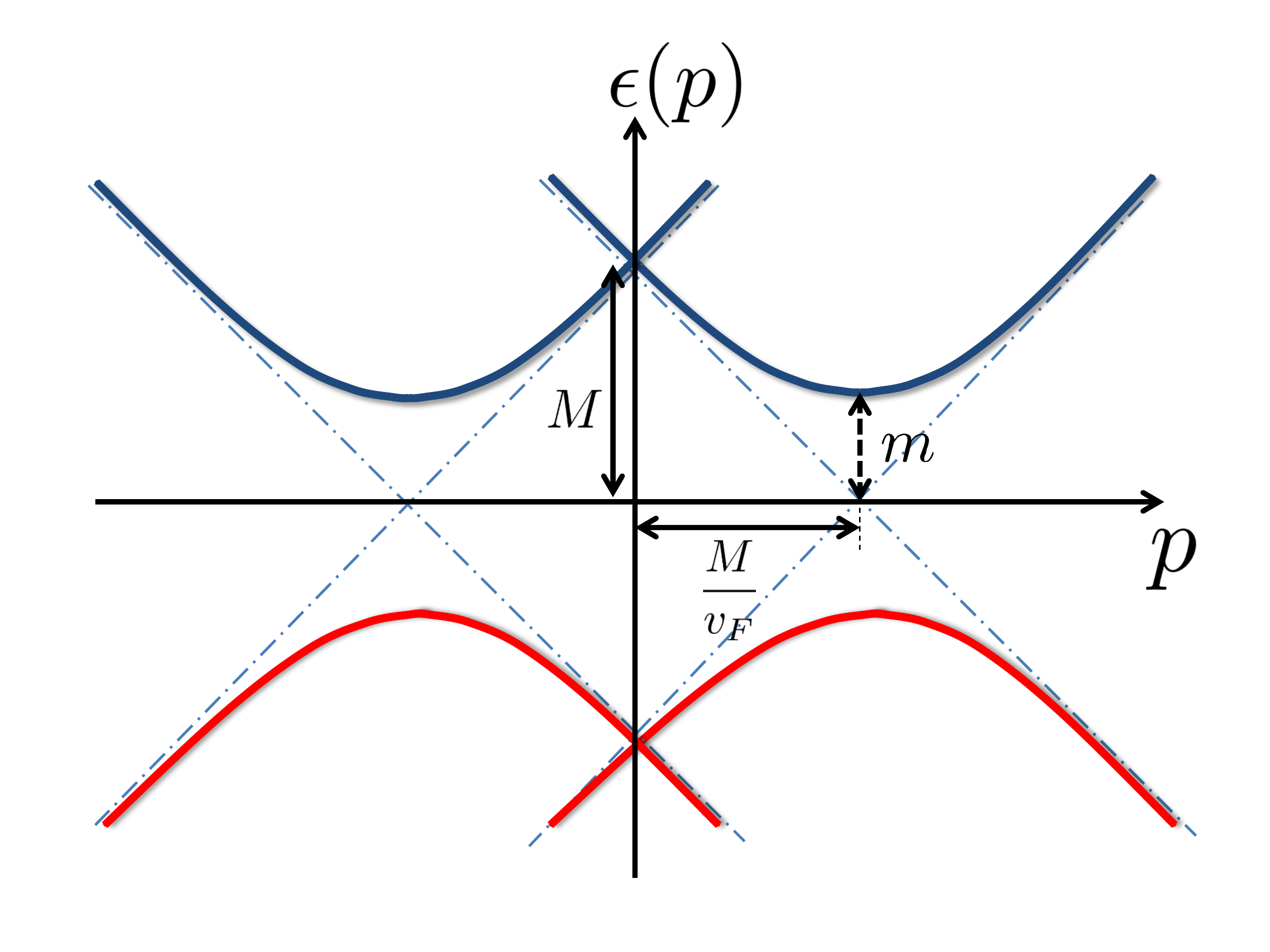}
\includegraphics[scale=.13]{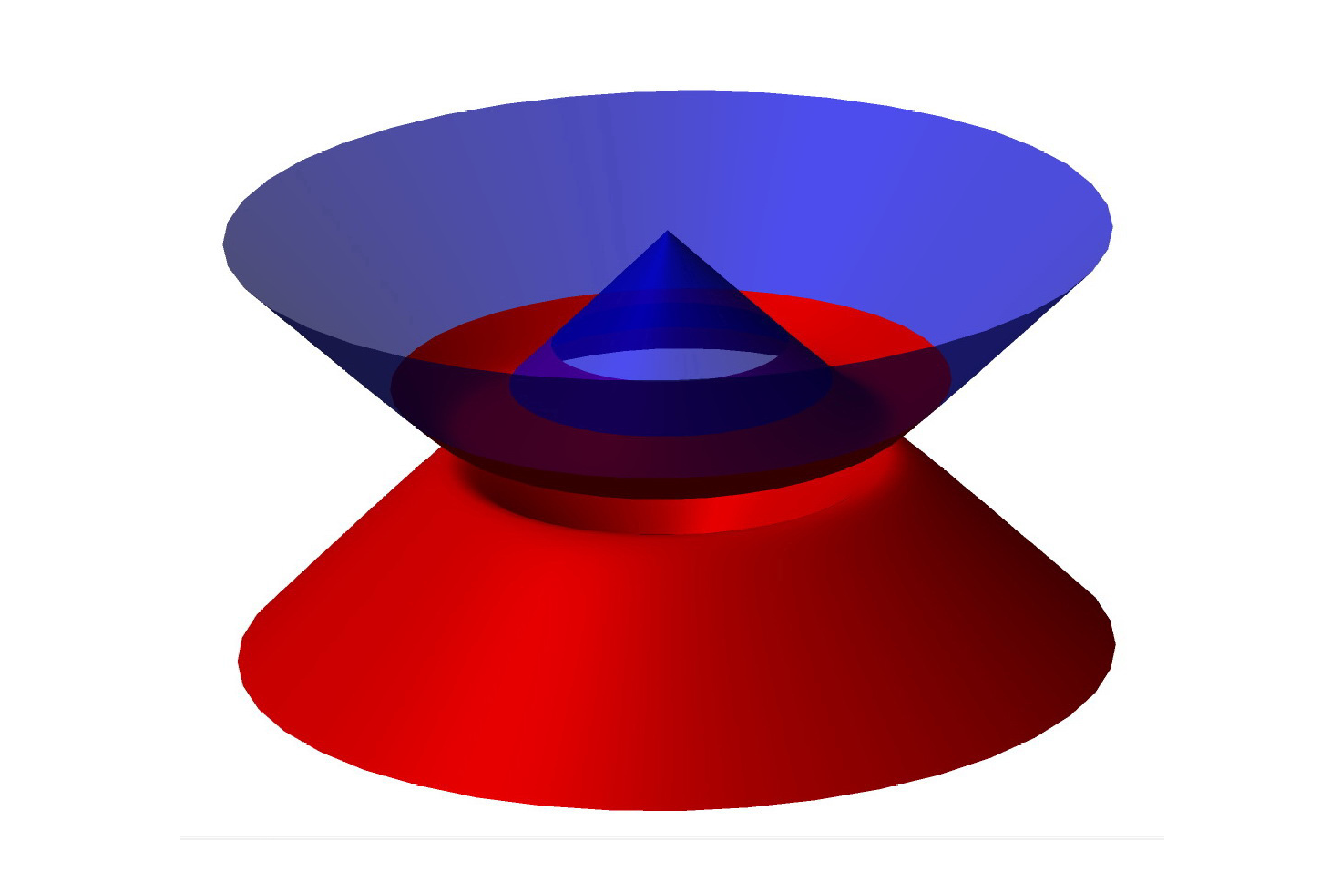}
\caption{(Color online) Left) The bandstructure of a Dirac ring system $\pm \epsilon(p)$ given by Eq. \ref{energy}, with asymptotic canonical velocity (slope) given by $v_F\hat k$. Right) 3D view of the same band structure, with the axis of rotation the energy axis and the plane of rotation the $p_x,p_y$ plane. The occupied (red) band is separated from the partially occupied (blue) band by a mass gap $2m$. They can be obtained by gapping out two intersecting Dirac cone with energy difference of $2M$.  A ring of filled states (red) at chemical potential $\mu>m$ lies in the ring of minima on the valence (blue) band.  }
\label{fig:bandstructure}
\end{figure}

%Before discussing the response of real Dirac-ring type materials, we first study the response of \emph{ideal} Dirac rings, where the dispersion takes the form
We now specialize to the Hamiltonians with a Dirac ring, which is the focus of this work. The energy dispersion of a Dirac ring system takes the form (Fig. \ref{fig:bandstructure})
\begin{equation}
\epsilon(p)=\pm \sqrt{m^2+(v_Fp-M)^2}
\label{energy}
\end{equation}
with $v_F$ the Fermi velocity, $m$ the (half) gap and $\frac{M}{v_F}$ the ring radius. It is known as a Dirac ring system because it is rotationally invariant about $\theta=\tan^{-1}\frac{p_y}{p_x}$, and has a ring of band minima with linear dispersion at radius $p=\frac{M}{v_F}$. Note that it reduces to the Dirac cone in Graphene when $M=0$, which was systematically studied in Ref. [19].  In TI heterostructure realization reported and analyzed in Refs. [23-26], $m$ is the mass gap induced by interlayer coupling and $M$ measures the extent of inversion symmetry breaking.

\subsubsection{Nonlinear response of Ideal Dirac rings}
We shall first study the proptotypical case of \emph{ideal} Dirac rings, where the gap $m$ and temperature $T$ are both zero. In the limit of small chemical potential $\mu\ll M $, which can be tuned by varying the gate voltage$^{26}$ , the filled states on the valence band form a very thin ring bounded by inner and outer Fermi momenta $p_I=\frac{M-\mu}{v_F}$ and $p_F=\frac{M+\mu}{v_F} $, i.e. with radial thickness $2\mu/v_F$. Being such a thin ring, its total current is thus well-approximated by $\vec j(M/v_F)$ in Eq. \ref{jp}, which is simple enough to visualize schematically and exactly evaluate analytically (Eq. \ref{jp1}).

The canonical velocity is simply given by
\begin{equation}
\vec v(p)|_{m=0}=\nabla_{\vec p}\epsilon(p)=v_F \text{sgn}(v_Fp-M)\hat p
\end{equation}
which is a vector of \emph{constant} magnitude $v_F$. It is always pointing in the radial direction, and is positive(negative) outside(inside) the Dirac ring. Such a scenario occurs when the filled states forms a non simply-connected 'ring-like' shape instead of a 'blob-like' shape. In this case, the lower dimensionality of the ring may allow for some novel kind of 'destructive interference' to occur between the contributions on both sides of the ring, and hence lead to negative differential resistance. Before proceeding with the calculations, let us attempt to understand that more intuitively.

\begin{figure}[H]
\includegraphics[scale=.43]{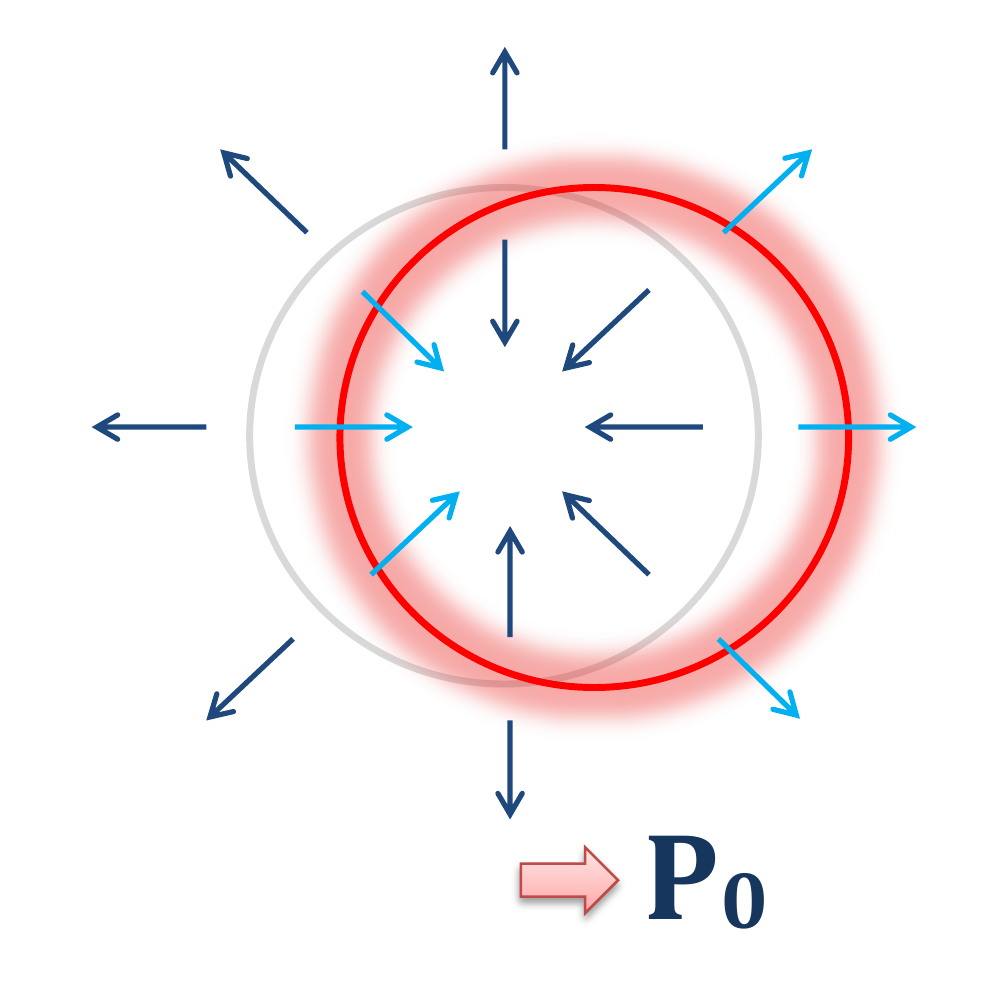}
\includegraphics[scale=.43]{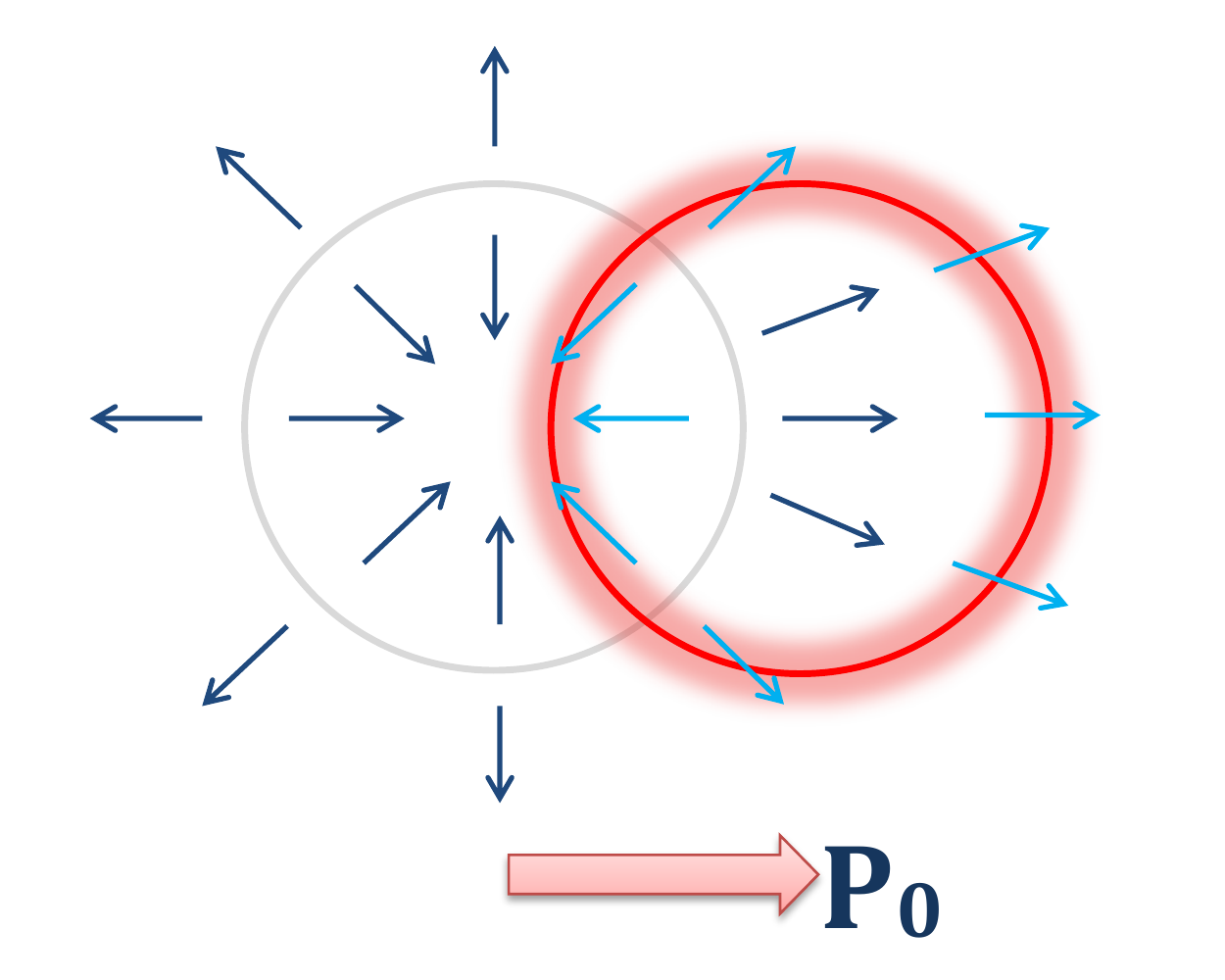}
\includegraphics[scale=.43]{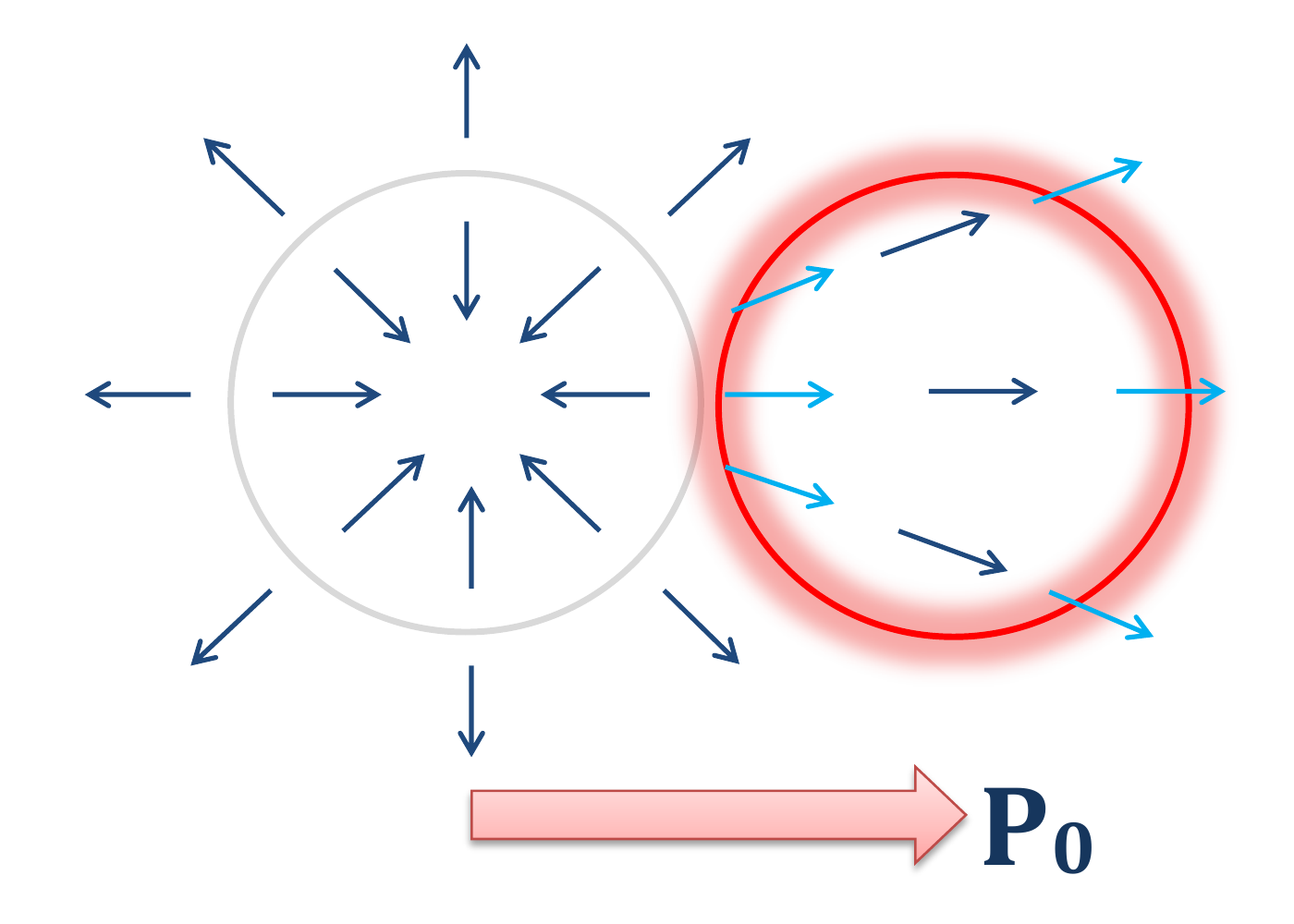}
\caption{(Color online) Left) For small driving impulse $p_0$, the current $\vec J= e \langle \vec v\rangle $ is relatively large as it is contributed by $\vec v$ whose horizontal components add constructively (light blue). Middle) As $p_0$ increases to the range $M<v_F p_0< 2M$, the horizontal components of $\vec v$ experience a partial cancellation (light blue), thereby lowering the resultant $J=|\vec J|$. Right) Beyond $p_0>2M/v_F$, the ring of filled states disentangles with the ring of Dirac nodes (grey). The contributions from $\vec v$ (light blue) add even more strongly, resulting in an even greater $J$. Very importantly, the behavior described here still holds for a ring thickened by (not too large) chemical potential or temperature, and even remain qualitatively true in the presence nonuniformities in $|\vec v|$, i.e. from nonzero mass.     }
\label{fig:interference}
\end{figure}

%Suppose we have very low chemical potential $\mu$, so that the occupied states form a very thin ring in p-space.
Without a driving field, $\vec p_0=\vec 0$ and the ring of filled states lie exactly above the ring of Dirac nodes. On either side of it, the vector field $\vec v(p)$ points in equal and opposite directions, thereby resulting in a zero net current in Eq. \ref{J}. and \ref{jp}. Upon a small impulse $p_0=|\vec p_0|$ from the driving field, the ring of filled states will be slightly displaced in reciprocal (momentum) space, leading to an imbalance between the contributions of $\vec (p)$ from inside and outside the ring. As illustrated in Fig. \ref{fig:interference}, $\vec J$ is relatively large for small $p_0$ because its arises from $\vec v$ contributions that point in the \emph{same} horizontal direction. As $|\vec p_0|$ increases to more than half the radius of the ring, contributions from inside the ring \emph{oppose} those from outside the ring, thereby leading to a \emph{decrease} in $|\vec J|$. Finally, for $ p_0$ larger than the radius of the ring, the filled states disentangle from the Dirac ring completely, and the $\vec v$ contributions point in \emph{same} horizontal direction again, adding up more strongly than before. As $p_0$ continue to increase, $\vec v$ will eventually become parallel, leading to a maximal current $|\vec J|$ proportional to the size of the ring.

\begin{figure}[H]
\includegraphics[scale=1.25]{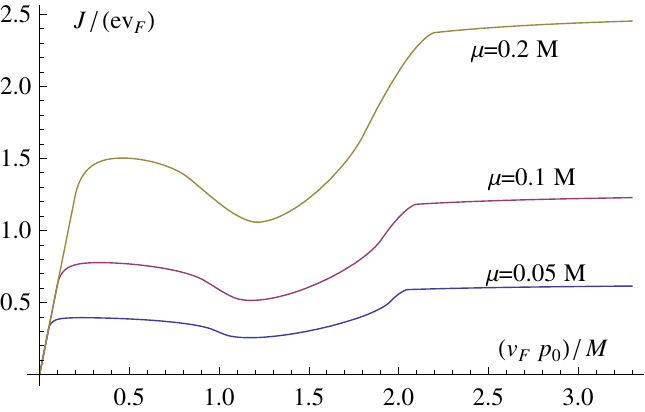}
\includegraphics[scale=1.25]{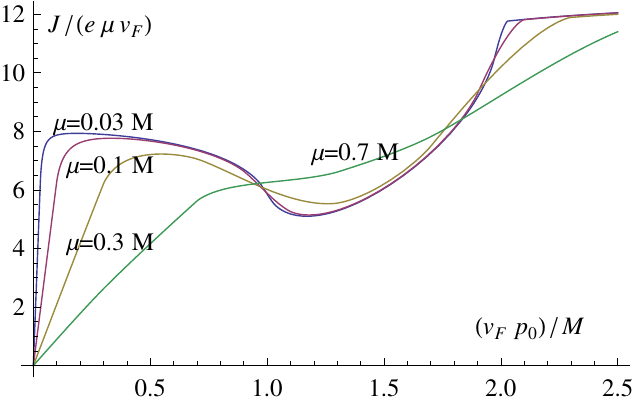}
\caption{(Color online) Plots of the current in Eq. \ref{Jmu0} against $v_Fp_0/M$ (defined as $Q^{-1}$ in Eq. \ref{jp2}), where $p_0$ is magnitude of the effective driving impulse. $p_0=eE_0\tau$ in the DC scattering limit, where $\tau$ is the scattering time. Left) Behavior of $J/(ev_F)$ for $\mu=0.05,0.1$ and $0.2$ in units of $M$. We see a strong linear regime at very small $p_0$, and a region of negative differential resistance $\frac{dJ}{dp_0}<0$ at larger values of $p_0$ due to destructively interfering $\vec v$. Right) The same plot normalized by $\mu$: $J/(\mu ev_F)$ for $\mu=0.03,0.1,0.3$ and $0.7$ in units of $M$. We now clearly see the asymptotic properties of the varies curves. As $\mu$ decreases, the response curve becomes 'sharper' due to the decreasingly thickness of the ring of filled states, and eventually converges nonuniformly to $G(Q,Q)$ (from Eq. \ref{jp2}) as $\mu\rightarrow 0$.  }
\label{fig:Jring}
\end{figure}

The above arguments suggest a response curve $J=|\vec J|$ that rises sharply to a moderately large value when $p_0$ is very small, decreases when $M<v_F p_0< 2M$, and increases to an even larger value for larger $p_0$. We identify the decreasing region as the region of negative differential resistance $\propto \frac{dJ}{dp_0}<0$. This agrees exactly with analytic expression derived in section \ref{gapless_analytic} and plotted in Fig. \ref{fig:Jring}:
%\red{Please decide how much to discuss about this...}
%There are nice analytic expressions for $j(p)$ in the zero $m$ and large $m$ limits.
\begin{eqnarray}
\vec J_{m=0,\mu\ll M}&=&4Q[(1+Q)\left( \text{EllipticE}\left[\alpha\right]+ \text{EllipticE}\left[\frac{\pi -\varphi }{2} ,\alpha\right]- \text{EllipticE}\left[\frac{\pi +\varphi }{2} ,\alpha\right]\right)\notag\\
&&+(1-Q)\left( \text{EllipticF}\left[\frac{\pi -\varphi}{2} ,\alpha\right]-\text{EllipticF}\left[\frac{\pi +\varphi}{2} ,\alpha\right]+ \text{EllipticF}\left[\frac{\pi}{2},\alpha\right]\right)\notag] \mu e \vec p_0\\
\label{Jmu0}
\end{eqnarray}
with $Q=\frac{M}{v_F p_0}$, $\alpha=\frac{4Q}{(1+Q)^2}$, $\varphi=\cos^{-1}\frac1{2Q}$, $\text{EllipticE}[\phi,\lambda]=\int_0^{\phi}\sqrt{1-\lambda \sin^2\theta}d\theta$ and $\text{EllipticF}[\phi,\lambda]=\int_0^{\phi}\frac1{\sqrt{1-\lambda \sin^2\theta}}d\theta$.  Note that only the dimensionless combination $Q=\frac{M}{v_F p_0}$ enters the expression.

\subsubsection{Distortion of a sinusoidal signal}

%As previously mentioned, the most common technique for producing low-power electromagnetic radiation at frequencies beyond 0.3 THz involve GaAs Schottky barrier diode upconverters which are inefficient in conversion beyond the first few harmonics$^{1-3}$....
 %Our Dirac ring, provides a decent upconversion until m=xx, making it a very promising candidate for frequency multiplication applications.

\begin{figure}[H]
\includegraphics[scale=.75]{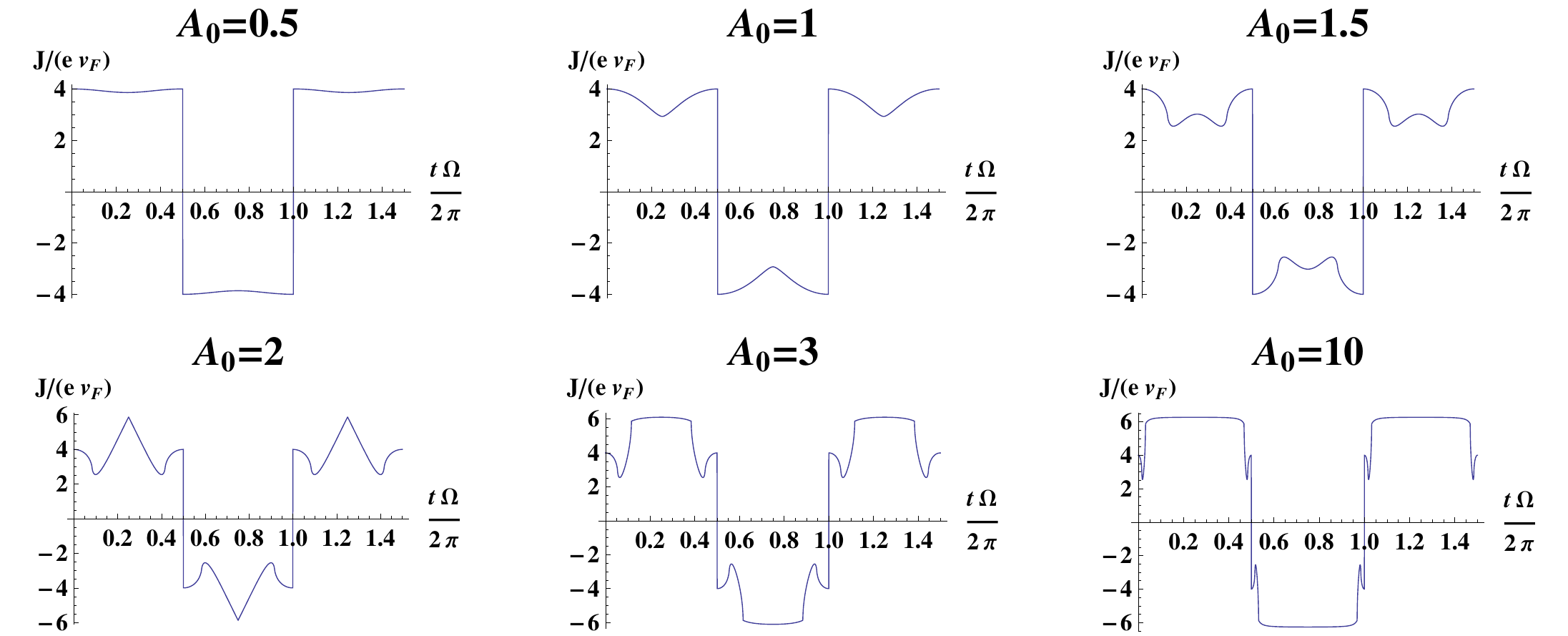}
\caption{(Color online) Output currents corresponding to input signals $E_{eff}=A_0\frac{\Omega M}{ev_F}\cos \Omega t$, where $A_0=0.5,1,1.5,2,3,10$ for very small $\mu$. Note the almost square-wave-like waveform for small and large $A_0$, where the response curve is mostly flat (Fig. \ref{fig:Jring}). The negative differential resistance part of the response curve produces the interesting lobes in the output current. They will be smoothed out at larger $\mu$, nonzero gap or nonzero temperature. In the $A_0\rightarrow \infty$ limit, the Dirac ring becomes irrelevant and the response curve reduce to that of Graphene ($\propto \text{sgn }p_0$) in the same limit.}
\label{fig:ringsignal}
\end{figure}

Below, we show how an ideal Dirac ring system distorts periodic signals of different amplitudes. Due to the segment of negative differential resistance in the response curves in Fig. \ref{fig:Jring}, additional kinks and lobes are introduced in the output signal $\vec J$. These lead to larger high-frequency components than what can be obtained with Graphene$^{19}$, whose distorted output signal is a square-wave which also appears in the $A_0\rightarrow \infty$ limit of the Dirac ring (Fig. \ref{fig:ringsignal}).

The extent of the lobes in the output can be quantified by the Fourier coefficients of the output current 
\begin{equation}
f_n=\frac1{\pi}\int_0^{2\pi} J(t)\sin nt
\end{equation}
Assuming an input impulse $\propto \sin t$, we find that only odd coefficients $f_{2n-1}$ are nonzero. The distribution of $\frac{|f_{2n-1}|}{|f_1|}$ characterizes the frequency multiplication efficacy, which is an important concern in the production of low-power electromagnetic radiation from lower frequencies. This is of particular exigence in the frequency window of 0.3 to 20 THz (commonly known as the Terahertz gap) where, despite a multitude of applications across engineering, material science and medical disciplines$^{19,27-29}$, inexpensive and compact sources for the THz radiation are still lacking. 

\begin{figure}[H]
\includegraphics[scale=.6]{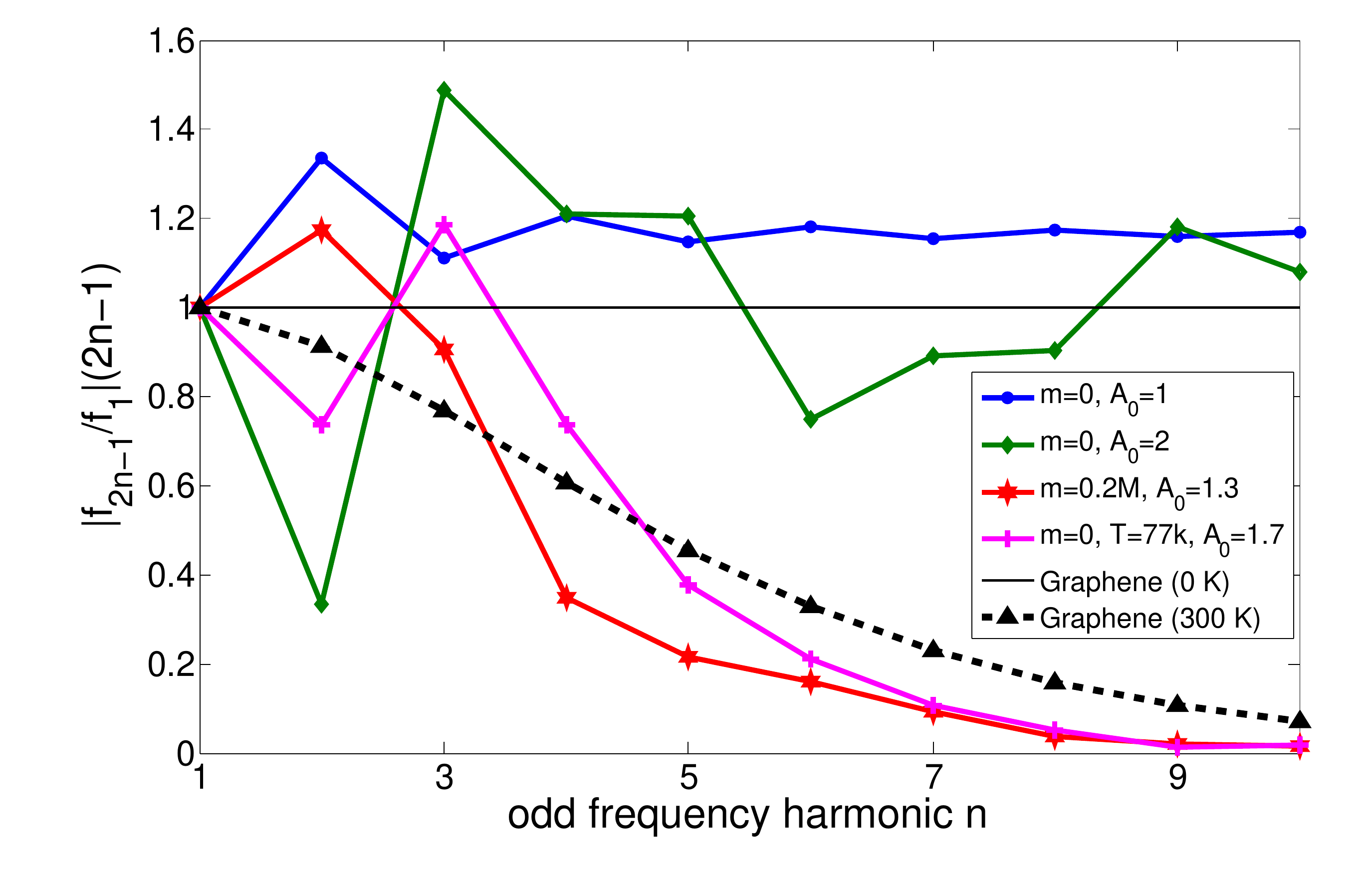}
\caption{(Color online) Ratios of the frequency multiplication factor $\frac{|f_{2n-1}|}{|f_1|}$ of the Dirac ring with that of zero-temperature Graphene ( $\propto \frac1{2n-1}$). The amplitudes $A_0$ are given by $E_{eff}=A_0\frac{\Omega M}{ev_F}\cos \Omega t$, which give maximal effective impulses $p_0^{max}=\frac{A_0M}{v_F}$. For reference, the continuous black line corresponds to Graphene at zero temperature $T$, which has the maximum possible frequency multiplication factor without negative differential resistance. It is exceeded by that of the gapless, zero-temperature Dirac ring at several different harmonics. This is expected from the appearance of additional, higher-frequency lobes in Fig. \ref{fig:ringsignal}, where the sharper lobes in the $A_0=2$ curve now manifest themselves as a suppression of the first higher harmonic $f_3$ relative to the $A_0=1$ case. At nonzero $T$ or $m$, the higher harmonics are exponentially suppressed. However, the $n=2$ or $n=3$ multiplication factor can still exceed the maximal possible without negative differential resistance (black line), as well as that of Graphene at $T=300 K,\mu=0$ under the same input signal as given in the discussion section.  }
\label{fig:fourier}
\end{figure}

It is interesting to compare the decay of $\frac{|f_{2n-1}|}{|f_1|}$ for the Dirac ring with the slowest decay spectrum possible \emph{without} negative differential resistance $\frac{dJ}{dp_0}<0$. The latter is given by a response curve proportional to $\text{sgn }p_0$, which can be realized$^{19}$ in Graphene at low (or zero) chemical potential and temperature relative to $v_Fp_0$. Simple computation yields a harmonic decay profile $|f^{graphene}_{2n-1}|\propto \frac1{2n-1}$, $n=1,2,...$. This is compared with that of the Dirac ring for various values of mass gap, temperature and input signal amplitudes in Fig. \ref{fig:fourier}. We see a larger frequency multiplication factor, i.e. larger $\frac{|f_{2n-1}|}{|f_1|}$ from the gapless Dirac ring across most of the harmonics $f_3,f_5,...,f_{19}$. When an energy scale is introduced by either nonzero gap or temperature, $f_{2n-1}$ decays exponentially. Even then, the frequency multiplication factor still outperforms that of Graphene for the the first few harmonics.

%The window of electromagnetic frequencies from 0.3 to 20 THz (also known as Terahertz gap) has been a very active area of research across a multitude of disciplines in engineering, material science and medicine$^{7,21-23}$. But despite their importance, inexpensive and compact sources for THz radiation are still lacking. Currently, low-power electromagnetic radiation within the Terahertz gap are most commonly produced via nonlinear multiplication (upconversion) of lower-frequency radiation$^{7,23-24}$. This, as often achieved with GaAs Schottky barrier diodes, have high upconversion efficiency for the first few harmonics but poor efficiencies for the higher order ones$^{7,23-24}$. In the following, we shall show that the electromagnetic response of Dirac ring systems provide a novel way for frequency upconversion that is efficient for high order harmonics. 

\subsection{Response of imperfect Dirac ring materials}

So far, we have considered Dirac ring systems with vanishing mass gap $m$, such as those realized in ${\text{Bi}}_{2}{\text{Se}}_{3}$ TI thin films heterostructures thicker than 6 quintuple layers (QLs)$^{23}$. To demonstrate the robustness of the regime of negative differential resistance, we shall now analyze Dirac ring systems which possess a gap and correspondingly a departure from perfect linear dispersion. They arise in sufficiently thin TI films where hybridization of the surface states on either side of the TI opens up a gap due to wavefunction overlap$^{24-25}$. This had been predicted$^{24-25}$ to occur and was indeed observed$^{23}$ in ${\text{Bi}}_{2}{\text{Se}}_{3}$ heterostructures thinner than 6 QLs. 

Real materials under experimental conditions furthermore experience non-negligible temperature effects and impurity scattering, both of which can undermine the preceding Dirac ring interference analysis. But very importantly, we shall show that the qualitative features of the response curve, particularly the region of negative differential resistance, remain robust. As long as the occupied states still occupy a ring-like region in reciprocal space, we indeed observe:
\begin{enumerate}
\item A rapidly increasing response $J$ for small $p_0$;
\item A region of decreasing $J$ $\left(\frac{dJ}{dp_0}<0\right)$ at moderate $p_0\approx M/v_F$, the radius of the ring;
\item An increasing, even larger $J$ for larger $p_0$.
\end{enumerate}
To understand why, it is useful to examine Fig. \ref{fig:interference} again. The non-monotonicity of the response (Center) is a consequence of the destructive interference of the contributions of $\vec v$ from inside and outside the ring. This is a \emph{generic} feature for a ring of energy minima, since the $\vec v= \nabla_{\vec p} \epsilon(p)$, which is always of opposite sides of the ring. In particular, note that it does not depend on the dispersion being linear, although the destructive interference will be less pronounced when the ring is thickened by large $\mu$ or $m$, or fuzzied by nonzero temperature $T$.

Below, we shall substantiate the above schematic arguments with detailed analyses and numerical results.

%\subsubsection{High chemical potential $\mu\geq M$, $m=0$}
%In this case, $\mu>\epsilon(p=0)=M$, and we have
%\begin{equation}
%\vec J_{m=0,\mu>M}=e\left ( \int_0^{p_F}+\int_0^{|p_I|}\right)\vec j(p)|_{m=0}dp=ep_0^2v_F \left ( \int_0^{P_F}+\int_0^{|P_I|}\right)G\left(P,\frac{M}{v_Fp_0}\right)dP\;\hat p_0
%\end{equation}

%There are a few interesting properties:
%\item For $p_0\gg p_F+\frac{M}{v_F}$, the whole region integrated is at one side of a Dirac 'cone', and $\vec J \rightarrow ev_F\pi p_F^2$, the same saturation value as that of Graphene from Mikhailov 2008.
%\begin{figure}[H]
%\includegraphics[scale=1.1]{J_mu_large.pdf}
%\caption{(Color online) Plots of $\frac{|J|}{ev_F(m^2+M^2)}$ vs. $v_Fp_0/M$ for $m=0$, for $\mu=1,2,4$ in units of $M$ (blue, purple, yellow). As we increase $\mu$, the kink due to the appearance of the lower Dirac cone at the center straightens out, and the current saturates to that like in a Dirac cone. }
%\label{fig:Jm0}
%\end{figure}

\subsubsection{Effect of nonzero band gap $m$}
\begin{figure}[H]
\includegraphics[scale=.6]{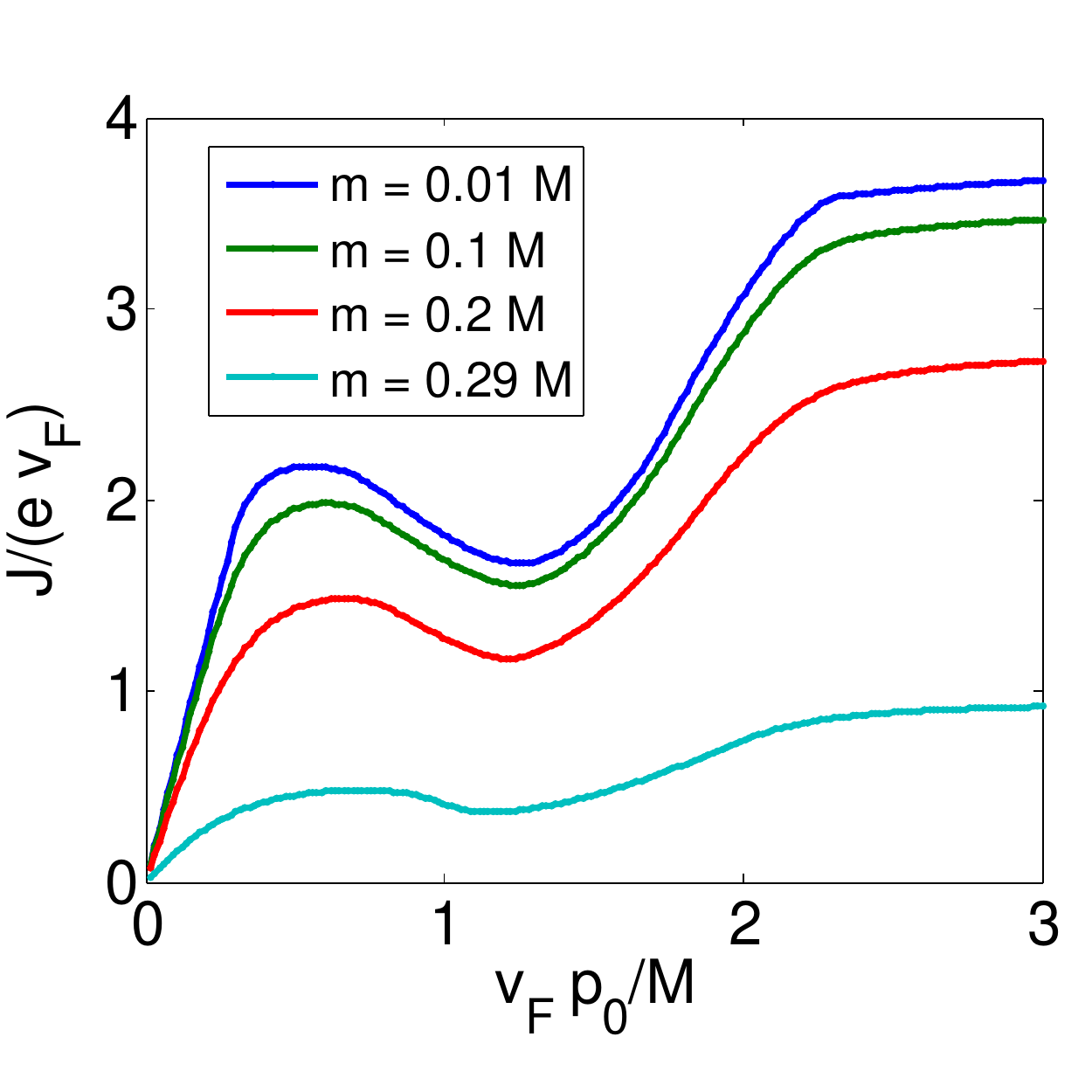}
\includegraphics[scale=.6]{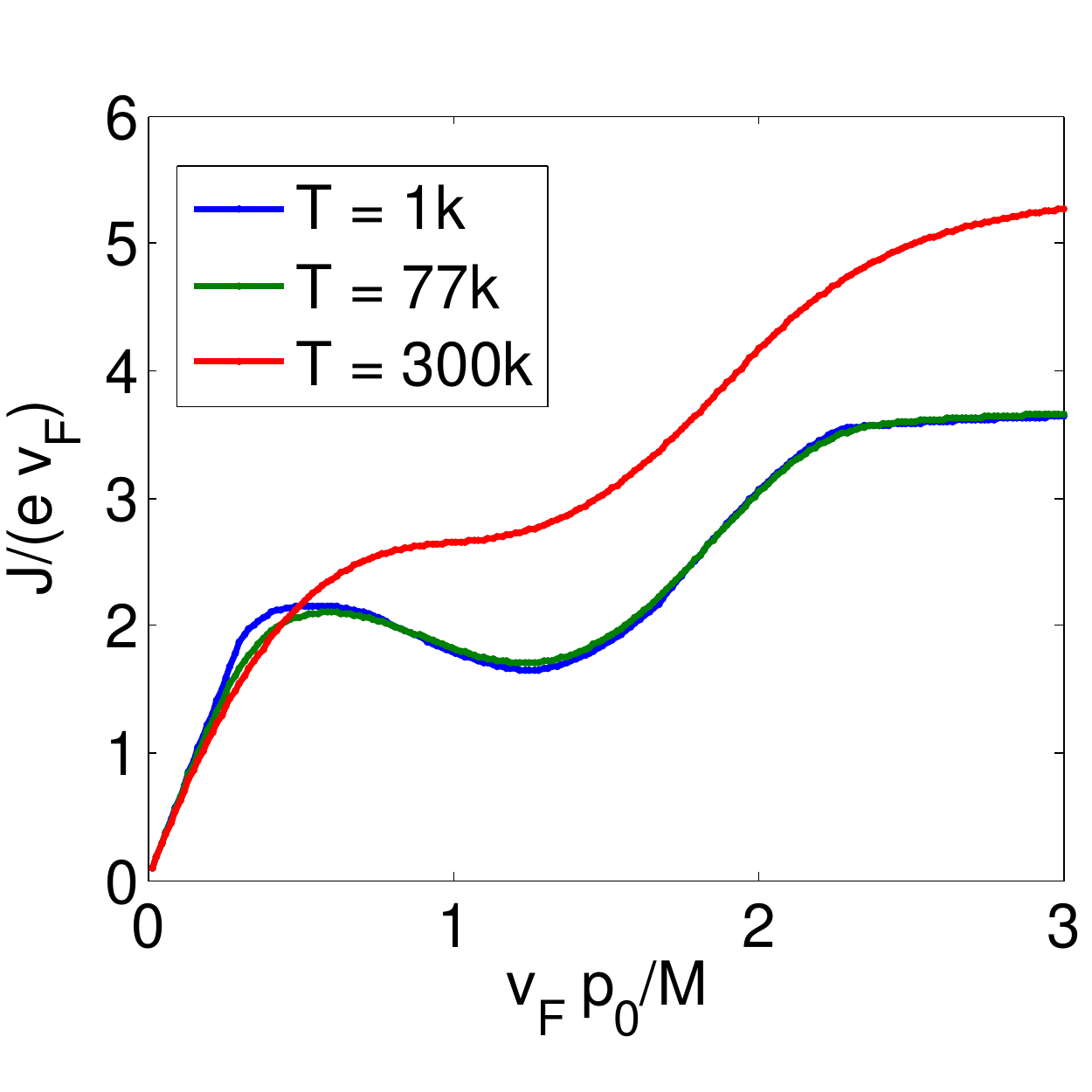}
\caption{(Color online) Left) The current response for different values of masses $m$, all at chemical potential $\mu=0.3M$ and temperature $T=0$. As expected, we observe the same qualitative, though 'rounder', response curve with increasing $m$. The current diminishes to zero as $m\rightarrow \mu$, when both $|\vec v|$ and the thickness of the filled ring diminishes. Right) The current response for different temperatures, with $M$ set to $0.1 eV$ and $\mu=0.3M$. There is essentially no difference in the curves at $T=1K$ and $T=77K$ (liquid nitrogen temperature).  }
\label{fig:mT}
\end{figure}

In a gapped Dirac ring system, Eq. \ref{energy} gives the canonical velocity in a gapped Dirac ring system as
\begin{equation}
\vec v(\vec p)=\nabla_{\vec p}\epsilon(\vec p)=\frac{v_Fp-M}{\sqrt{m^2+(v_Fp-M)^2}}v_F\hat p
\label{energy2}
\end{equation}
which vanishes linearly near the Dirac ring. As such, we expect a more 'rounded' response curve, as shown in Fig. \ref{fig:mT} (Left). Note that Eq. \ref{energy2} also describes the small $p_0$ response in a Rashba system. The latter, however, has a asymptotically quadratic dispersion which leads to a (rather uninteresting) linear response for large $p_0> m/v_F$, which also suppresses the region of negative differential resistance. 

The results for nonzero mass gap $m$ are shown in Fig. \ref{fig:mT} (Left). The response curves exhibit the same qualitative shape, but with a broader linear response regime for small $\frac{v_Fp_0}{M}$, which is studied in more detail in Sect. \ref{analytic_gapped}. Examples of such cases include the bulk states of HgTe$^{31}$ and GaAs$^{32}$ quantum wells with inversion symmetry breaking.

\subsubsection{Effects of nonzero temperature $T$}

In the presence of nonzero temperature, the Dirac ring is smeared out over a radius of $\Delta p\approx T/v_F$. Furthermore, hole carrier from the $-\epsilon(\vec p)$ band also participate in current transport. The smearing of the Dirac ring results in qualitatively similar modifications to the response curve as increasing $\mu$ or $m$. However, the participation of the hole carriers undermines the destructive interference of $\vec v$ described in Fig. \ref{fig:interference}, and can destroy the characteristic non-monotonicity and thus negative differential resistance of the response curve at sufficiently high $T$. Nonetheless, we numerically find (Fig. \ref{fig:mT} (Right)) that the ideal response curve retains its qualitative shape till \emph{room} temperature ($300$K) for $M\approx 0.1eV$, the inversion symmetry breaking scale in experimental realizations. This cutoff temperature is approximately $3000M$ in units of $K/eV$.

%\subsubsection{Frequency multiplication \red{Let's remove it for this version?}}

\subsubsection{Effect of scattering}
\label{sect:boltzmann}

Scattering is inevitable in real materials. For instance, a scattering
time of $\tau=4.2\times10^{-14}$s was reported for a Bi-based TI compound$^{33}$. Here, we shall model the effect of scattering by a classical relaxation time
$\tau$, and show that its effects can be incorporated into our
calculations by replacing the input driving field $\vec E$ with a
'renormalized' \emph{effective} driving field $\vec E_{eff}$.

The electronic states are distributed according to a time-dependent occupation function $g(p(t),t)$ that obeys the Boltzmann equation
\begin{equation}
\frac{dg}{dt}=\frac{\partial g}{\partial t}+ \frac{\partial g}{\partial \vec p}\cdot\frac{d\vec p}{dt}=\frac{f-g}{\tau}
\label{boltzmann0}
\end{equation}

where $f=f(\vec p(t))=F_0(\vec p(t))=(1+e^{\beta(\epsilon_{\vec p}-\mu)})^{-1}$ is the local equilibrium electron distribution, $F_0$ being the Fermi-Dirac distribution which can deviate from $g=g(p(t),t)$,the true, non-equilibrium distribution. Being the equilibrium distribution, the functional form of $f(\vec p)$ remains unchanged, and all fluctuations in time occurs only through the time-varying momentum of an electron $\vec p(t)$. By contrast, $g$ can deviate from this equilibrium through its explicit time dependence, although it will relax towards $f$ with a characteristic time $\tau$. We will assume spatial homogeneity throughout.

Eq. \ref{boltzmann0} has an explicit solution though elementary calculus:
\begin{equation}
\frac{dg(t)}{dt}=\int^t_{-\infty}dt' \frac{e^{(t'-t)/\tau}}{\tau}\frac{df(t')}{dt'}
\label{dg}
\end{equation}
where the dependence on $\vec p(t)$ has been made implicit. Intuitively, $\frac{dg}{dt}$ is a Laplace transform of $\frac{df}{dt}$, with contributions from earlier times exponentially suppressed due to scattering. As derived in Sect. \ref{sect:boltzmann2}, Eq. \ref{dg} has an explicit solution:

\begin{equation}
g(\vec p,t)=F_0\left(\vec p + e\int^t \vec E_{eff}(t')dt'\right)
\end{equation}
where
\begin{eqnarray}
\vec E_{eff}(t)&=& \int^tdt'e^{(t'-t)/\tau}\frac{d\vec E(t')}{dt}
\label{Eeff0}
\end{eqnarray}
is the \emph{effective} electric field that contains the effect of scattering.
%$\vec p_0=\vec p(0)=\vec p_d(0)$ is the initial (unperturbed) momentum of a particle, and $G$ an arbitrary distribution that depends on the initial conditions of the disturbance from equilibrium. The $G$ term always satisfies $\frac{dG}{dt}=0$, so that $\frac{dg}{dt}=\frac{dF_0}{dt}=\frac{df}{dt}$. The $G(\vec p_0)$ constant was determined by letting $|\vec E|\rightarrow 0$, so that $g$ approaches the unperturbed distribution $g(\vec p_0)$. Of course, $G(\vec p_0)$, being a constant, will not contribute to electron transport.
Essentially, $\vec E_{eff}$ is contributed by past \emph{increments} of the electric field $ \vec E$ that are exponentially suppressed with a characteristic time $\tau$. As $\tau$ increases, the increments have more time to constructively interfere. Physically, $|\vec E_{eff}|$ increases as $\tau$ increases because a longer ballistic motion contributes greater to the momentum shift of each particle. Of course, $|\vec E_{eff}(t)|\leq \text{max}_{t'<t}|\vec E(t')|$. The manifest causality structure results from the time-reversal asymmetry of the system cf. Eq. \ref{boltzmann0}.

For a simple sinusoidal driving field  $\vec E(t)= \vec E_0 \sin \Omega t$, Eq. \ref{Eeff0} (or Eq. \ref{J2}) can be solved for the \emph{exact} effective field
\begin{equation}
\vec E_{eff}(t)= \vec E_0 \frac{\Omega \tau}{1+(\Omega \tau)^2}\left(\cos \Omega t + \tau \Omega \sin \Omega t\right)
\label{Eeffsin}
\end{equation}
leading to the effective impulse
\begin{eqnarray}
\vec p&\rightarrow& \vec p-  \frac{e \tau\vec E_0}{1+(\Omega \tau)^2}\left(\sin \Omega t - \tau \Omega \cos \Omega t\right)\notag\\
&=& \vec p-  \frac{e \tau\vec E_0}{\sqrt{1+(\Omega \tau)^2}}\sin(\Omega t - \tan^{-1}\Omega\tau)%notag\\
%&\approx& \vec p-  \frac{e \tau\vec E_0}{\sqrt{1+(\Omega \tau)^2}}\sin\left(\Omega (t-\tau)+\frac{\Omega^2\tau^2}{3}\right)
\label{Eeffsin2}
\end{eqnarray}
In other words, scattering leads to a renormalization of the amplitude by a factor $\frac{\Omega\tau}{\sqrt{1+(\Omega \tau)^2}}$ and a phase retardation corresponding to the duration $\frac{\tan^{-1}\Omega \tau}{\Omega}\approx \tau\left(1+ \frac{\Omega^2\tau^2}{3}\right)$. The latter is trivial for a single frequency mode, but will lead to interference effects when there is a mixture of modes. That will further studied in Sect. \ref{sect:boltzmann2}, in the limits of weak and strong scattering. Note that an expression very similar to Eq. \ref{Eeffsin2} also appeared in Ref. [19] in the context of radiation damping. But here $\tau$ has a more generic interpretation, encompassing all generic damping mechanisms.

The DC limit can be recovered by taking $\Omega t -\frac{\pi}{2} \ll 1$, so that $\vec E(t)\approx \vec E_0$. The effective impulse is equal to $e\vec E_0 t$ in the ballistic regime, but saturates at $e\vec E_0 \tau$ when $t$ is increased to $\tau$ and effect of scattering is felt.

\section{Discussion}

Through an approach based on the Boltzmann equation, we have analytically and numerically obtained the novel nonlinear response curve of a Dirac ring system representing Topological Insulator thin films. Due to the special topology of a ring-shaped region of filled states, the response curve is intrinisically non-monotonic, leading to a regime with negative differential resistance which heightens frequency upconversion of a periodic signal. This property remains robust in real, experimentally fabricated samples in the terahertz gap, with nonzero mass gap, chemical potential, temperature and impurity scattering rate.

One test of our results will be to reproduce the output signals in Fig. \ref{fig:ringsignal}. To produce the $A_0=1$ curve, for instance, we can input either an electrical or optical sinusoidal signal$^{34}$ of amplitude $E_0=
\frac{\sqrt{1+(\Omega \tau)^2}}{\Omega \tau}\frac{\Omega M}{e
v_F}A_0$ on a sample of ${\text{Bi}}_{2}{\text{Se}}_{3}$ thin film thicker than
6 QLs. Experimentally obtained parameters $M\approx 0.1eV$,
$v_F=4.52\times10^5m/s$,$^{23}$ and $\tau=4.2\times10^{-14}s$,$^{33}$ correspond to a realistic electric field of magnitude$^{34}$ $E_0=5.27\times 10^4 V/cm$ and frequency$^{35}$ $\Omega=100GHz$. The output current should closely agree with that in Fig. \ref{fig:ringsignal} at the temperature of  liquid nitrogen $T=77K$, and deviate slightly from it at room temperature $T=300K$ (cf. Fig. \ref{fig:mT}(Right)). All other output signal shapes in Fig. \ref{fig:ringsignal} may be reproduced by varying the input signal amplitude. Since the effective impulse from the field increases linearly with the scattering time $\tau$, TIs with much larger $\tau$ will exhibit the aforesaid negative differential resistance and nonlinear response behavior at much weaker physical electric fields. This will be very attractive for low power applications, and is well on track to becoming a reality with the rapid development of TI material growth techniques as well as the discovery of novel TI materials. 

\section{Methods}

\subsection{The current response of a gapless Dirac ring}
\label{gapless_analytic}
Here we derive the analytic expression for the response of an (ideal) gapless Dirac ring, for which the $\mu\rightarrow 0$ result in Eq. \ref{Jmu0} is a special case.

Substituting the canonical velocity $\vec v(p)|_{m=0}=\nabla_{\vec p}\epsilon(p)=v_F \text{sgn}(v_Fp-M)\hat p$ in Eq. \ref{jp}, where $F_0(\epsilon)=\theta(-\epsilon)$ at temperature $T=0$, we obtain%where $j(p)$ has a closed form expression. Later, we will show that the nonlinear response remains qualitatively similar when $0<m<M$ and $0<k_BT<M$. We have
\begin{eqnarray}
 \vec j(p)|_{m=0} &=& v_FpF_0(\epsilon(p))\int\text{Sgn}\left(v_F|\vec p +\vec p_0|-M\right)\frac{\vec p + \vec p_0}{|\vec p + \vec p_0|} d\theta \notag\\
&=&\hat p_0 v_FF_0(\epsilon(p))\int\text{Sgn}\left(\sqrt{p^2+p_0^2+2pp_0\cos\theta}-M/v_F\right)\frac{p(p_0+ p\cos\theta)}{\sqrt{p_0^2+p^2+2pp_0\cos\theta}} d\theta \notag\\
%&=& v_FF_0(\epsilon(p))\text{Sgn}(v_Fp-M)\left(2\text{sgn}[P-1]P\left((P-1)E\left[\frac{-4P}{(P-1)^2}\right]-(P+1)K\left[\frac{-4P}{(P-1)^2}\right]\right)\right)\vec p_0\notag\\
&=&v_FF_0(\epsilon(p))G\left(\frac{p}{p_0},\frac{M}{v_Fp_0}\right)\vec p_0
\label{jp1}
\end{eqnarray}
where $G(P,Q)$, $P=\frac{p}{p_0}$, $Q=\frac{M}{v_Fp_0}$ is a closed-form expression given by
\begin{eqnarray}
\frac{G(P,Q)}{2P}&=&(1+P)\left( \text{EllipticE}\left[\alpha\right]+ \text{EllipticE}\left[\frac{\pi -\varphi }{2} ,\alpha\right]- \text{EllipticE}\left[\frac{\pi +\varphi }{2} ,\alpha\right]\right)\notag\\
&&+(1-P)\left( \text{EllipticF}\left[\frac{\pi -\varphi}{2} ,\alpha\right]-\text{EllipticF}\left[\frac{\pi +\varphi}{2} ,\alpha\right]+ \text{EllipticF}\left[\frac{\pi}{2},\alpha\right]\right)\notag\\
%\frac{G(P,Q)}{2P}&=&(1+P) \text{EllipticE}\left[\frac{4 P}{(1+P)^2}\right]+(1+P) \text{EllipticE}\left[\frac{1}{2} \left(\pi -\text{Re}\left[\text{ArcCos}\left[\frac{1+P^2-Q^2}{2 P}\right]\right]\right),\frac{4 P}{(1+P)^2}\right]\notag\\
%&& -(1+P) \text{EllipticE}\left[\frac{1}{2} \left(\pi +\text{Re}\left[\text{ArcCos}\left[\frac{1+P^2-Q^2}{2 P}\right]\right]\right),\frac{4 P}{(1+P)^2}\right]\notag\\
%&&+(1-P) \text{EllipticF}\left[\frac{1}{2} \left(\pi -\text{Re}\left[\text{ArcCos}\left[\frac{1+P^2-Q^2}{2 P}\right]\right]\right),\frac{4 P}{(1+P)^2}\right]\notag\\
%&&+(P-1)\text{EllipticF}\left[\frac{1}{2} \left(\pi +\text{Re}\left[\text{ArcCos}\left[\frac{1+P^2-Q^2}{2 P}\right]\right]\right),\frac{4 P}{(1+P)^2}\right]-(P-1) \text{EllipticF}\left[\frac{\pi}{2},\frac{4 P}{(1+P)^2}\right]\notag\\
\label{jp2}
\end{eqnarray}
with $\alpha=\frac{4P}{(1+P)^2}$, $\varphi=\text{Re}\left[\cos^{-1}\left[\frac{1+P^2-Q^2}{2 P}\right]\right]$, $\text{EllipticE}[\phi,\lambda]=\int_0^{\phi}\sqrt{1-\lambda \sin^2\theta}d\theta$ and $\text{EllipticF}[\phi,\lambda]=\int_0^{\phi}\frac1{\sqrt{1-\lambda \sin^2\theta}}d\theta$.
%where $P=\frac{p}{p_0}$ is the magnitude of $\vec p$ relative to that of $\vec p_0$
The functional dependence $G(P,Q)=G\left(\frac{p}{p_0},\frac{M}{v_Fp_0}\right)$ implies that of the 3 parameters $p,p_0$ and $M/v_F$, only two can affect the result independently. For instance, the effect of letting $M\rightarrow 2M$ is equivalent to that of $p_0\rightarrow p_0/2$, $p\rightarrow p/2$ and $G\rightarrow 2G$.
 %We see that apart from the Fermi factor, $\vec j\propto \vec p_0$ up to $F(P,Q)$ that depends only on $P=\frac{p}{p_0}$ and $Q=\frac{M}{p_0v_F}$.

%In the limit of $M\propto Q=0$, we recover the case for Graphene:
%\begin{equation}
%G(P,0)=2\text{sgn}[P-1]P\left((P-1)EllipticE\left[\frac{\pi}{2},\frac{-4P}{(P-1)^2}\right]-(P+1)EllipticF\left[\frac{\pi}{2},\frac{-4P}{(P-1)^2}\right]\right)
%\end{equation}
We are now ready to calculate $\vec J$ proper. It is just a sum over annuli of different radii $p$, each contributing a current $\vec j(p)$. From Eq. \ref{energy}, $\mu$ defines two Fermi surfaces with inner and outer Fermi momenta $p_I=\frac{M-\mu}{v_F}$ and $ p_F=\frac{M+\mu}{v_F}$. Considering the interesting case of small $\mu<M$\footnote{If $\mu\geq M$, the occupied momenta lie in a disk of radius $p_F$ for the lower band, and of radius $|p_I|$ for the upper band. Note that all incompletely filled bands must be included. Of course, the model must be lattice regularized if we want to study huge $p_F$ of the order of $\pi$.}, the occupied momenta lie in a ring of inner and outer radii $p_I$ and $p_F$. In the limit of $\mu\rightarrow 0$, the occupied states form a very thin annulus and an illuminating closed-form expression exists for $\vec J$:
%\red{Move this sentence elsewhere: In general, $J$ is proportional to  $\frac{eM^2}{v_F}$ times a function of the ratios $\frac{p_0}{M/v_F}$ and $\frac{\mu}{M/v_F}$.}
\begin{equation}
\vec J_{m=0,\mu\ll M}=e\int_{\frac{M-\mu}{v_F}}^{\frac{M+\mu}{v_F}}  \vec j(p)|_{m=0}dp\approx_{\mu\rightarrow 0} 2\mu ep_0v_F G\left(\frac{M}{v_Fp_0},\frac{M}{v_Fp_0}\right)\hat p_0=2\mu e  G(Q,Q)\vec p_0
\end{equation}
which is just Eq. \ref{Jmu0}. %The ring geometry exhibits strong nesting properties, and encourages very strong response even at small perturbations. The occupied states form a thin ring of at radius $\approx M/v_F$, and the response is very non-linear, even non-monotonic, as can be seen from its limiting behavior at $\mu=0$ (writing $Q=\frac{M}{v_F p_0}$):
%\begin{eqnarray}
%\frac{G(Q,Q)}{2Q}&=&(1+Q) \text{EllipticE}\left[\frac{4 Q}{(1+Q)^2}\right]+(1+Q) \text{EllipticE}\left[\frac{1}{2} \left(\pi -\text{ArcCos}\left[\frac{1}{2 Q}\right]\right),\frac{4 Q}{(1+Q)^2}\right]\notag\\
%&&-(1+Q) \text{EllipticE}\left[\frac{1}{2} \left(\pi +\text{ArcCos}\left[\frac{1}{2 Q}\right]\right),\frac{4 Q}{(1+Q)^2}\right]\notag\\
%&&-(Q-1) \text{EllipticF}\left[\frac{1}{2} \left(\pi -\text{ArcCos}\left[\frac{1}{2 Q}\right]\right),\frac{4 Q}{(1+Q)^2}\right]\notag\\
%&&+(Q-1)\text{EllipticF}\left[\frac{1}{2} \left(\pi +\text{ArcCos}\left[\frac{1}{2 Q}\right]\right),\frac{4 Q}{(1+Q)^2}\right]-(Q-1) \text{EllipticF}\left[\frac{\pi}{2},\frac{4 Q}{(1+Q)^2}\right]\notag
%\label{GQQ}
%\end{eqnarray}
For larger $\mu$, numerical integration yields the plots in Fig. \ref{fig:Jring} which has the following properties:
\begin{itemize}
\item The current is notably proportional to $v_F$. This property is unique to the quasi-1D shape of the ring. While the velocity operator $\vec v$ is proportional to $v_F$, the area of the ring is independent of it. This is because its radius is proportional to it, while its thickness is inversely proportional to it.
\item The response at any finite $\mu$ does not converge uniformly to Eq. \ref{Jmu0}: At small perturbations $p_0$, the ring feels opposite velocity fields at both sides of the Fermi surface (FS), and there is a resultant linear regime. From Eq. \ref{Jmu0}, one can show that
\begin{equation}
 J_{m=0,\mu<M,p_0<\frac{\mu}{v_F}}\approx  2\pi eM p_0
\end{equation}

for $0<p_0<\frac{\mu}{M}$ to a very high degree of accuracy, as shown in Fig. \ref{fig:Jring}. Next comes a nonlinear regime where the response is negative approximately in the range $\mu<0.6 v_F$.

\item At $p_0=\frac{M}{v_F}$ or $Q=1$, we pass a special point where the ring untangles from the ring of Dirac nodes. Here,
\begin{equation}
 J_{m=0,\mu<M,p_0}=\frac{2\pi e\mu M}{v_F}
\end{equation}

\item For larger $p_0\gg \frac{\mu}{v_F}$ or $Q\gg 1$, we have
\begin{equation}
 J_{m=0,\mu\leq M}=\frac{4\pi e\mu M}{v_F}
\end{equation}
It can also be shown that
\begin{equation}
 J_{m=0,\mu\geq M}=\frac{2\pi e(\mu^2+M^2)}{v_F}
\end{equation}
\end{itemize}

\subsection{Current response of a gapped Dirac ring}
\label{analytic_gapped}

Here we derive some analytic results for the response due to small perturbations about a thin gapped Dirac ring, so as to understand how the ring structure affect the linear response of an otherwise massive system.

For sufficiently small $\mu-m$ and $p_0$, the inequality $m\gg |v_F|p+p_0|-M|$ holds and the canonical velocity is approximately %, i.e. the regime where $m\gg \mu + v_F p_0 + \sqrt{\mu(\mu_2v_F p_0)}$,
\begin{equation} \vec v(\vec p)\approx \frac{v_F}{m}(v_F \vec p -M \hat p)\end{equation}
This is also what we have in the vicinity of the ring of minima of a system with Rashba splitting. Substituting it into Eq. \ref{jp} like before, we obtain
\begin{equation}
\vec j(p)|_{m\text{ large}}\approx \frac{v_FF_0(\epsilon(p))}{m}\left(2\pi p v_F-MG(p/p_0,0)\right)\vec p_0
\end{equation}

The first term is linear in $\vec p_0$, agreeing with that of usual materials with quadratic dispersion $\propto \frac{m}{2}p^2$; upon doing the $p$ integral, we recover
\begin{equation}
\vec J_{quadratic}=\frac{e\pi v_F^2 p_F^2}{m}\vec p_0
\end{equation}
where $p_F$ is the (outer) Fermi momentum. The second term is a Dirac-like contribution that modifies the response from the lowest-order quadratic approximation. Integrating both terms, we obtain
\begin{equation}
 \vec J|_{m\text{ large}}\approx\frac{ev_F^2}{m}(p_F^2- p_I^2)\vec p_0 - \frac{M}{m}\left(\vec J_{Dirac}(p_F)- \vec J_{Dirac}(p_I)\right)
\end{equation}
where $ p_F,P_I=\frac{M\pm\sqrt{\mu^2-m^2}}{v_F}$ and
$\vec J_{Dirac}(p)=ep_0^2v_F\int_0^PG(P,0)dP\; \hat p_0$ denotes the Dirac cone ($M=0$) current in Eq. \ref{jp} and \ref{jp2}.

\subsection{The effective driving field due to scattering}
\label{sect:boltzmann2}
Here we show that the non-equilibrium state distribution $g$ is of the form
\begin{equation}
g(\vec p,t)=F_0\left(\vec p +e\int^t \vec E_{eff}(t')dt'\right),
\end{equation}
and derive $\vec E_{eff}$ in terms of the original driving field $\vec E$. The Boltzmann Eq. \ref{boltzmann0} has an explicit solution Eq. \ref{dg}:
\begin{equation}
\frac{dg(t)}{dt}=\int^t_{-\infty}dt' \frac{e^{(t'-t)/\tau}}{\tau}\frac{df(t')}{dt'}
\end{equation}
where $f=f(\vec p(t))=F_0(\vec p(t))$. To motivate the solution to Eq. \ref{dg}, we first solve it for a periodic driving field. For each fourier component $\Omega$, we have $\frac{d\vec p}{dt}=-e \vec E(t)=-e \vec E_\Omega e^{i\Omega t}$, so that

\begin{eqnarray}
\frac{dg(t)}{dt}&=&\int^t_{-\infty}dt' \frac{e^{(t'-t)/\tau}}{\tau}\frac{df}{d\epsilon}\frac{d\epsilon}{d\vec p}\cdot \frac{d\vec p}{dt}\notag\\
&=&-e\int^t_{-\infty}dt' \frac{e^{(t'-t)/\tau}}{\tau}\frac{dF_0(\epsilon(\vec p))}{d\epsilon}\vec v(\vec p)\cdot \vec E_\Omega e^{i\Omega t'}\notag\\
&=&-e\frac{dF_0(\epsilon(\vec p))}{d\epsilon}\vec v(\vec p)\cdot \vec E_\Omega \int^t_{-\infty}dt' \frac{e^{(t'-t)/\tau+i\Omega t'}}{\tau}\notag\\
&=&-e\frac{dF_0(\epsilon(\vec p))}{d\epsilon}\vec v(\vec p)\cdot \vec E_\Omega e^{i\Omega t}\int_0^{\infty}dT\frac{e^{-(i\Omega +1/\tau)T}}{\tau}\notag\\
&=&-e\frac{dF_0(\epsilon(\vec p))}{d\epsilon}\vec v(\vec p)\cdot \vec E_\Omega \frac{e^{i\Omega t}}{ 1+i\Omega \tau}\notag\\
&=&-e\frac{dF_0(\epsilon(\vec p))}{d\epsilon}\frac{\vec v(\vec p)\cdot \vec E(t)}{ 1+i\Omega \tau}
\label{dg2}
\end{eqnarray}
For a generic periodic driving field, Eq. \ref{dg2} still holds, but with a sum over all fourier modes $\Omega$.  Hence the effect of the $\tau$ damping is the replacement of the fourier coefficients $ (E_d)_\Omega = \frac{E_\Omega}{1+i\Omega \tau} $, which can also be guessed from elementary considerations.

As such, let's define a \emph{damped} momentum $\vec p_d$ which responds to the damped electric field $\vec  E_d$ with fourier coefficients $ (E_d)_\Omega$. From Eq. \ref{dg2}, we obviously have\footnote{Care has to be taken in handling these PDEs: while $g$ depends on $t$ both explicitly and implicitly through $\vec p(t)$, the dependence of $f$ on $t$ is only implicit through $\vec p$ and $\vec p_d$, respectively before and after integrating out the damping effect.}

\begin{equation}
\frac{dg(\vec p(t),t)}{dt}=\frac{df(\vec p_d(t))}{dt}=-e\nabla f\cdot \vec E_d
\label{dg3}
\end{equation}
This integrates to
\begin{equation}
%g(\vec p,t)=F_0\left(\vec p_0-e\int_0^t\vec E_d(t')dt'\right)+G\left(\vec p+e\int_0^t E(t')dt'\right)-G(\vec p_0)
g(\vec p,t) = F_0\left(\vec p+e\int^t(\vec E_d(t')-\vec E(t'))dt'\right)=F_0\left(\vec p+e\int^t\vec E_{eff}(t')dt'\right)
\label{g}
\end{equation}
where
\begin{eqnarray}
\vec E_{eff}(t)&=& \frac{1}{2\pi}\int dt' \vec E(t')\int  \frac{e^{i\Omega(t-t')}d\Omega}{1+\frac{1}{i\Omega \tau}}\notag\\
&=&\int^t \vec E(t')\frac{e^{(t'-t)/\tau}}{\tau} dt'-\vec E(t)\notag\\
&=& \int^tdt'e^{(t'-t)/\tau}\frac{d\vec E(t')}{dt}
\label{Eeff}
\end{eqnarray}

\subsubsection{The diffusion limit of small $\tau$}
In the limit of strong scattering, only recent memories of $\Delta \vec E$ survive, and an expansion about $t'=t$ in Eq. \ref{Eeff} gives
\begin{eqnarray}
\vec E_{eff}(t)&\approx &-\tau \frac{d \vec E(t)}{dt}-\tau ^2\frac{d^2 \vec E(t)}{d^2t}-\tau^3 \frac{d^3 \vec E(t)}{dt^3}-...\notag\\
&=& \frac{\tau \frac{d}{dt}}{\tau \frac{d}{dt}-1}\vec E(t)
\label{smalltau}
\end{eqnarray}
The above expansion is valid in the regime $\tau < \Delta t$, where $\Delta t\sim \Omega^{-1}$ is the characteristic time scale at which $\vec E$ varies. %It is important to note that in this short $\tau$ regime, $\vec E_{eff}$ is proportional to the \emph{time gradients} of $\vec E$, and not $\vec E$ itself.
To linear order, the current is thus
\begin{eqnarray}
\vec J_{linear}& =& -e\int d^2p \vec v(\vec p) F_0\left(\vec p+ e\int^t \vec E_{eff}(t')dt'\right)\notag\\
&\approx& -e\int d^2p  \left(F_0(\vec p) + e\frac{dF_0(\vec p)}{d\epsilon_{\vec p}}\vec v\cdot \int^t \vec E_{eff}(t')dt'\right)\vec v\notag\\
&\approx & -e^2 \int d^2p \frac{dF_0(\vec p)}{d\epsilon_{\vec p}}\vec v\cdot \left(\tau \vec E(t) + \tau^2 \frac{d \vec E(t)}{dt} +...\right)\vec v
\label{J}
\end{eqnarray}
This linear approximation coincides exactly with the usual derivation of the Drude formula. If we allow for large perturbations, we will instead have
\begin{eqnarray}
\vec J_{non-linear}& =& -e\int d^2p \vec v(\vec p) F_0\left(\vec p+e \tau\vec E(t)+e\tau^2\frac{d\vec E(t)}{dt}-...\right)
\label{J2}
\end{eqnarray}

Clearly, the argument of $F_0$ is the unperturbed crystal momentum $\vec p$ plus the impulse from $\vec E$ over an effective duration of $\tau$. Note that it is \emph{not} a Taylor expansion of $\vec E(t-\tau)$; it is the series form of the exponentially suppressed field given in Eq. \ref{Eeff}, with terms given by $\tau^n \frac{d^{n-1}\vec E(t)}{dt^{n-1}}$.

\subsubsection{The ballistic limit of large $\tau$}
First, we consider what happens when $\tau\rightarrow \infty$, which is the limit studied in Ref. [$19$]. Since the system is driven periodically, $\vec E(t')$ has equal positive and negative contributions, and
\begin{equation}
\lim_{\tau\rightarrow \infty}\frac{dg(t)}{dt}=\left\langle\frac{df}{dt}\right\rangle =0
\label{largetau}
\end{equation}
That $\frac{dg}{dt}=0$ forces the $g$ to have the functional form $F_0(\vec p - e \vec A(t))$, where explicit time dependence only enters through $\vec A(t)=-\int^t \vec E(t')dt'$. $F_0$ is fixed to be the Fermi-Dirac distribution by considering the limit $|\vec E|\rightarrow 0$. In a sense, this argument is a simple yet insightful semiclassical justification of the minimal substitution $\vec p \rightarrow \vec p-e\vec A$ for a distribution: only through this substitution will $g$ remain a constant as we follow a particle, as it should be in the absence of any other force.

Now, let us consider first-order scattering contributions $\sim\frac{1}{\tau}$. From the second line of Eq. \ref{Eeff}, we have
\begin{equation}
g(\vec p ,t )\approx F_0\left(\vec p - e\int^t \vec E(t')dt'+\frac{1}{\tau}\int^t \vec E(t')e^{(t'-t)/\tau}\right)
\label{largetau2}
\end{equation}
The $\frac1{\tau}$ term keeps track of the effects of scattering. Note that is a Laplace transformation in $\vec E$ itself, and not of $\frac{d\vec E}{dt}$ like in the diffusive limit. %The physical differences

\subsection*{References and Notes}
\begin{itemize}
\item[1.]
Qi, X. L. \& Zhang, S. C. The quantum spin Hall effect and
topological insulators. {\it Phys. Today} {\bf 63,} 33-38 (2010).
\item[2.]
Qi, X. L. \& Zhang, S. C. Topological insulators and
superconductors. {\it Rev. Mod. Phys.} {\bf 83,} 1057 (2011).
\item[3.]
Hasan, M. Z. \& Kane, C. L. Colloquium: Topological insulators. {\it
Rev. Mod. Phys.} {\bf 82,} 3045 (2010).

\item[4.]
Bernevig, B. A. \& Hughes, T. L. \& Zhang, S. C. Quantum Spin Hall
Effect and Topological Phase Transition in HgTe Quantum Wells. {\it
Science} {\bf 314,} 1757-1761 (2006).
\item[5.]
K\"onig, M. et al. Quantum spin Hall insulator state in HgTe quantum wells. {\it
Science} {\bf 318,} 766-770 (2007).
\item[6.] Fu, L. \& Kane, C. L. \& Mele, E. J. Topological insulators in three dimensions. {\it Phys. Rev. Lett.} {\bf
98,} 106803  (2007).
\item[7.]
Hsieh, D. et al. A topological Dirac insulator in a quantum spin Hall phase. {\it
Nature} {\bf 452,} 970-974 (2008).
\item[8.]
Zhang, H.\&  Liu, C.-X. \& Qi, X.-L. \&  Dai, X. \&  Fang, Z. \&
Zhang, S.-C. Topological insulators in {Bi$_2$Se$_3$},
{Bi$_2$Te$_3$} and {Sb$_2$Te$_3$} with a single Dirac cone on the
surface. {\it Nat. Phys.} {\bf 5}, 438 (2009).
\item[9.]
Chen, Y. L. et al. Experimental realization of a three-dimensional topological insulator, ${\text{Bi}}_{2}{\text{Te}}_{3}$. {\it
Science} {\bf 325,} 178-181 (2009).
\item[10.]
Peng, H. et al. Aharonov-Bohm interference in topological insulator nanoribbons. {\it
Nature Materials} {\bf 9,}  225-229 (2010).

\item[11.]
Chadov, S. \& Qi, X. \& Kübler, J. \& Fecher, G. H. \& Felser, C. \& Zhang, S. C.  Tunable multifunctional topological insulators in ternary Heusler compounds. {\it
Nature Materials} {\bf 9,}  541-545 (2010).
\item[12.]
Zhang, X. \& Zhang, H. J. \& Wang, J. \& Felser, C. \& Zhang, S.-C.
Actinide Topological Insulator Materials with Strong Interaction.
{\it Science} {\bf 335,} 1464 (2012).
\item[13.]
Yan, B. \& Jansen, M. \& Felser, C.  A large-energy-gap oxide topological insulator based on the superconductor $\text{Ba}\text{Bi}{\text{O}}_{3}$. {\it
Nature  Physics} {\bf 9,}  709-711 (2013).

\item[14.]
Zhang, X. \&  Wang, J. \& Zhang, S.-C. Topological insulators for
high-performance terahertz to infrared applications. {\it Phys. Rev.
B} {\bf 82,} 245107 (2010).

\item[15.]
Novoselov, K. S. et al. Two-dimensional gas of massless Dirac fermions in
Graphene. {\it Nature} {\bf 438,} 197-200 (2005).
\item[16.]
Nair, R. R. et al. Fine Structure Constant Defines Visual Transparency of Graphene.
{\it Science} {\bf 320,} 1308-1308 (2008).
\item[17.]
Xia, F. \& Mueller, T. \&  Lin, Y.-m. \& Valdes-Garcia, A. \&
Avouris, P. Ultrafast Graphene photodetector. {\it Nature
Nanotechnology} {\bf 4,} 839-843 (2009).
\item[18.]
Castro Neto, A. H. \& Guinea, F. \& Peres, N. M. R. \& Novoselov, K.
S. \& Geim, A. K. The electronic properties of Graphene. {\it Rev.
Mod. Phys.} {\bf 81,} 109-162 (2009).
\item[19.]
Mikhailov, S. A. \& Ziegler. K, Nonlinear electromagnetic response
of Graphene: frequency multiplication and the self-consistent-field
effects. {\it J. Phys.: Condens. Matter} {\bf 20,} 384204 (2008).
\item[20.]
 Wright, A. R. \&  Xu, X. G. \& Cao, J. C. \&  Zhang,C. Strong nonlinear
optical response of Graphene in the terahertz regime. {\it Appl.
Phys. Lett.} {\bf 95,} 072101 (2009).
\item[21.] Hendry, E. \& Hale,
P. J. \& Moger, J. \&  Savchenko, A. K. \& Mikhailov, S. A. Coherent
Nonlinear Optical Response of Graphene. {\it Phys. Rev. Lett.} {\bf
105,} 097401  (2010).
\item[22.]
Gunn, J. B. Instability of current in III-V semiconductors. {\it IBM J. Res.tiDevelop.} {\bf 8,} 141-159 (1964).
\item[23.]
Zhang, Y. et al. Crossover of the three-dimensional topological
insulator ${\text{Bi}}_{2}{\text{Se}}_{3}$ to the two-dimensional
limit. {\it Nature Physics} {\bf 6,} 584-588 (2010).

\item[24.]
Liu, C. X. et al. Oscillatory
crossover from two-dimensional to three-dimensional topological
insulators. {\it Phys. Rev. B} {\bf 81,} 041307 (2010).
\item[25.]
Linder, J. \& Yokoyama, T. \& Sudb\o{}, A. Anomalous finite size
effects on surface states in the topological insulator
${\text{Bi}}_{2}{\text{Se}}_{3}$. {\it Phys. Rev. B} {\bf 80,}
205401 (2009).

\item[26.]
Wang, J. \& Mabuchi, H. \& Qi, X.-L. Calculation of divergent photon absorption in ultrathin films of a topological insulator. {\it Phys. Rev. B} {\bf 88,} 195127 (2013).

\item[27.]
Sherwin, M. S. \& Schmuttenmaer, C. A. \& Bucksbaum, P H (ed) 2004
Opportunities in THz science. {\it Report of a DOE-NSF-NIH Workshop
(Arlington, VA)}.
\item[28.]
Faist, J. \& Capasso, F. \& Sivco D. L. \& Sirtori, C. \&
Hutchinson, A. L. \& Cho, A. Y. Quantum cascade laser. {\it Science}
{\bf 264,} 553-556 (1994).
\item[29.]
Siegel, P. H. Terahertz technology. {\it IEEE Trans. Microw. Theory
Tech.} {\bf 50,} 910-928 (2002).
\item[30.]
Raisanen, A. V. Frequency multipliers for millimeter and
submillimeter wavelengths. {\it Proc. IEEE} {\bf 80,} 1842-1852 (1992).
\item[31.]
Klipstein, P. C. Structure of the quantum spin Hall states in HgTe/CdTe and InAs/GaSb/AlSb quantum wells. {\it Phys. Rev. B} {\bf 91,} 035310 (2015).
\item[32.]
Giglberger, S. et al. Rashba and Dresselhaus spin splittings in semiconductor quantum wells measured by spin photocurrents. {\it Phys. Rev. B} {\bf 75,} 035327 (2007).

\item[33.]
 Taskin, A. A.\& R. Zhi \&  S. Satoshi \& Segawa K.\& A. Yoichi
Observation of Dirac Holes and Electrons in a Topological Insulator.
{\it Phys. Rev. Lett.} {\bf 107,} 016801 (2011).
\item[34.]
 Paul, M. J. et al. High-field terahertz response of
Graphene. {\it New Journal of Physics} {\bf 15,} 085019 (2013).
\item[35.]
 Chen, S. Q. et al. Broadband optical and microwave nonlinear
response in topological insulator. {\it Optical Materials Express}
{\bf 4,} 597-586 (2014).
\end{itemize}

\section*{Additional Information}
\subsection{Author Contributions}
CH. Lee and X. Zhang contributed equally to the manuscript. B. Guan contributed to the calculations for Fig. 4 and 5. All authors reviewed the manuscript.

\subsection{Competing financial interests}
The author(s) declare no competing financial interests.

\section*{Acknowledgments}
We thank S. A. Mikhailov for helpful discussions. C.H. Lee is supported by a fellowship from the Agency of Science, Technology and Research of Singapore. X. Zhang is supported by the National Natural Science Foundation of China (No.11404413). 

\end{document}